\documentclass[english]{article}
\usepackage[T1]{fontenc}
\usepackage[latin9]{inputenc}
\usepackage{amsbsy}
\usepackage{amssymb}

\makeatletter
\date{}
\usepackage{mathrsfs}

\makeatother

\usepackage{babel}

\begin{document}

\title{A note on the short-time quantum propagator}

\author{{\normalsize Ali Sanayei}%
\thanks{Email: ali.sanayei@uni-tuebingen.de%
}\\
\emph{\small Institute for Theoretical Physics, University of T\"ubingen}\emph{
}\\
\emph{\small Auf der Morgenstelle 14, D-72076 T\"ubingen, Germany}\emph{}\\
}
\maketitle
\begin{abstract}
In the Feynman formalism of quantum mechanics one encounters a postulate,
namely, that the propagator in an infinitesimal time-interval is the
classical wave function. This postulate, which was later studied thoroughly
by Holland, was recently highlighted by using the improved Makri-Miller
propagator. The present note, whose conclusion is in agreement with
that recent achievement, demonstrates that the Heisenberg picture
of quantum mechanics invariably includes the Feynman postulate and
is able to yield a proof for it. In other words, by starting out from
the Heisenberg picture, it is proved that when terms of second-order
in time can be neglected, the dynamics of a system is classical.

\end{abstract}

\section{Introduction}

In the formulation of quantum mechanics from the Feynman point of
view, there exists a postulate reading that, in an infinitesimal time-interval,
the propagator is just the classical wave function {[}4,10{]}. It
was then discussed by Holland {[}9{]} that along each infinitesimal
interval, the motion of the particle is classical and that the full
quantum mechanics at later times will be recovered by the superposition
principle. Accordingly, it was recently shown by using an improved
version of the Makri-Miller propagator that when terms of $\mathcal{O}\left(\Delta t^{2}\right)$
can be neglected, the classical trajectories arise from a short-time
quantum propagator {[}12{]}. The same idea was also applied to generate
the full quantum mechanics of the cold Bose atoms around a crossing
of quantum waveguides with different geometries using the Chapman-Kolmogorov
identity {[}13{]}. Both the proof and its applications seem to be
crucial for the foundation of quantum mechanics because they shed
light on one of the famous Oxford Questions {[}11{]} which reads:
``Does the classical world emerge from the quantum, and if so which
concepts are needed to describe this emergence?''

It is commonly stated that the classical level arises when $h$ (Planck's
constant) tends to zero. This viewpoint can at first sight be realized
by a fundamental equation of quantum mechanics (as Dirac called it
{[}1{]}) which represents a mapping from the commutation of two dynamical
variables into their Poisson bracket; that is,

\begin{equation}
\left[u,v\right]=i\hslash\left\{ u,v\right\} ,\end{equation}
where $u$ and $v$ are arbitrary functions of a set of canonical
coordiantes and momenta {[}14{]}. This is because for a non-zero Poisson
bracket when $h\rightarrow0$, $u$ and $v$ commute and therefore
become classical dynamical variables. This approach, however, did
not satisfy Bohm and hence he reformulated the problem by means of
invoking different orders of time and also the concept of ``quantum
potential'' {[}15{]}.

The question to be addressed here is as follows: How does the Heisenberg
picture of quantum mechanics handle the Feynman postulate? That is
to say, can the Heisenberg picture present a proof for the Feynman
postulate, or should the statement in this picture be in the form
of an axiom? The purpose of the present note is to demonstrate that
it is actually possible to start out from the Heisenberg picture of
quantum mechanics and from there to prove the Feynman postulate. More
precisely, the following statement will be offered: The Heisenberg
picture inherently includes the notion that, when terms of $\mathcal{O}\left(\Delta t^{2}\right)$
can be neglected, the dynamics of a system is classical. The next
Section is dedicated to the proof.

\section{Heisenberg picture and Feynman postulate}

The Heisenberg picture of quantum mechanics supposes that the dynamical
variables are matrices and hence do not satisfy the commutative axiom
of multiplication. Those dynamical variables vary with time according
to Heisenberg's equation of motion

\begin{equation}
i\hslash\frac{d\hat{\xi}}{dt}=\left[\hat{\xi},\mathrm{\hat{H}}\right],\end{equation}
where $\hat{\xi}$ denotes a dynamical variable and $\mathrm{\hat{H}}$,
corresponding to the total energy in the Heisenberg picture, denotes
a linear operator which is just the transform of the Hamiltonian operator
occuring in the Schr\"odinger picture. The relation of the given dynamical
variable when written either in the Heisenberg or the Schr\"odinger
picture is given by the following unitary transformation {[}3,6{]}:

\begin{equation}
\hat{\xi}_{H}\left(t\right)=e^{i\mathrm{\hat{H}}\frac{\left(t-t_{0}\right)}{\hslash}}\hat{\xi}e^{-i\mathrm{\hat{H}}\frac{\left(t-t_{0}\right)}{\hslash}}\end{equation}

Given the position $\hat{\mathrm{q}}$ and momentum $\hat{\mathrm{p}}$
as two dynamical variables, one can thus by (3) find the corresponding
Heisenberg representations. Let $\left|\psi\left(t_{0}\right)\right\rangle $
denote the state of a dynamical system at time $t_{0}$ and define
the following representations for position and momentum in the Heisenberg
picture:

\begin{equation}
\left\langle \psi\left(t_{0}\right)\left|\hat{\mathrm{q}}_{H}\left(t\right)\right|\psi\left(t_{0}\right)\right\rangle \equiv Q\left(t\right)\end{equation}

\begin{equation}
\left\langle \psi\left(t_{0}\right)\left|\hat{\mathrm{p}}_{H}\left(t\right)\right|\psi\left(t_{0}\right)\right\rangle \equiv P\left(t\right)\end{equation}
Let $\Delta t=t-t_{0}$ denote a time interval. One can then with
the help of (4) and (5) and also using Hadamard's lemma {[}5,7{]}
write:

\[
Q\left(t_{0}+\Delta t\right)\]

\[
=\left\langle \psi\left(t_{0}\right)\left|\hat{\mathrm{q}}_{H}(t_{0})+\frac{i}{\hslash}\left[\mathrm{\hat{H}},\hat{\mathrm{q}}\right]\Delta t+\left(\frac{i}{\hslash}\right)^{2}\left[\mathrm{\hat{H}},\left[\mathrm{\hat{H}},\hat{\mathrm{q}}\right]\right]\frac{\Delta t^{2}}{2}+\mathcal{O}\left(\Delta t^{3}\right)\right|\psi\left(t_{0}\right)\right\rangle \]

\begin{equation}
=Q\left(t_{0}\right)+\left\langle \psi\left(t_{0}\right)\left|\frac{i}{\hslash}\left[\mathrm{\hat{H}},\hat{\mathrm{q}}\right]\right|\psi\left(t_{0}\right)\right\rangle \Delta t+\mathcal{O}\left(\Delta t^{2}\right);\end{equation}

\[
P\left(t_{0}+\Delta t\right)\]

\[
=\left\langle \psi\left(t_{0}\right)\left|\hat{\mathrm{p}}_{H}(t_{0})+\frac{i}{\hslash}\left[\mathrm{\hat{H}},\hat{\mathrm{p}}\right]\Delta t+\left(\frac{i}{\hslash}\right)^{2}\left[\mathrm{\hat{H}},\left[\mathrm{\hat{H}},\hat{\mathrm{p}}\right]\right]\frac{\Delta t^{2}}{2}+\mathcal{O}\left(\Delta t^{3}\right)\right|\psi\left(t_{0}\right)\right\rangle \]

\begin{equation}
=P\left(t_{0}\right)+\left\langle \psi\left(t_{0}\right)\left|\frac{i}{\hslash}\left[\mathrm{\hat{H}},\hat{\mathrm{p}}\right]\right|\psi\left(t_{0}\right)\right\rangle \Delta t+\mathcal{O}\left(\Delta t^{2}\right).\end{equation}
By assuming an infinitesimal time-interval, that is to say, $\Delta t\rightarrow0$,
and neglecting the terms of order $\mathcal{O}\left(\Delta t^{2}\right)$
in (6) and (7), these equations yield:

\begin{equation}
\lim_{\Delta t\rightarrow0}\frac{Q(t_{0}+\Delta t)-Q(t_{0})}{\Delta t}=\dot{Q}=\left\langle \psi\left(t_{0}\right)\left|\frac{i}{\hslash}\left[\mathrm{\hat{H}},\hat{\mathrm{q}}\right]\right|\psi\left(t_{0}\right)\right\rangle \end{equation}

\begin{equation}
\lim_{\Delta t\rightarrow0}\frac{P(t_{0}+\Delta t)-P(t_{0})}{\Delta t}=\dot{P}=\left\langle \psi\left(t_{0}\right)\left|\frac{i}{\hslash}\left[\mathrm{\hat{H}},\hat{\mathrm{p}}\right]\right|\psi\left(t_{0}\right)\right\rangle \end{equation}
One can with the help of Ehrenfest theorem {[}2{]} and the definitions
(4) and (5) write:

\begin{equation}
\dot{Q}=\frac{d}{dt}\left\langle \psi\left(t_{0}\right)\left|\hat{\mathrm{q}}_{H}(t)\right|\psi\left(t_{0}\right)\right\rangle =\frac{i}{\hslash}\left\langle \psi\left(t_{0}\right)\left|\left[\mathrm{\hat{H}},\hat{\mathrm{q}}_{H}\right]\right|\psi\left(t_{0}\right)\right\rangle \end{equation}

\begin{equation}
\dot{P}=\frac{d}{dt}\left\langle \psi\left(t_{0}\right)\left|\hat{\mathrm{p}}_{H}(t)\right|\psi\left(t_{0}\right)\right\rangle =\frac{i}{\hslash}\left\langle \psi\left(t_{0}\right)\left|\left[\mathrm{\hat{H}},\hat{\mathrm{p}}_{H}\right]\right|\psi\left(t_{0}\right)\right\rangle \end{equation}
The right-hand sides of (10) and (11) should respectively be equal
to the right-hand sides of (8) and (9); hence:

\begin{equation}
\frac{i}{\hslash}\left\langle \psi\left(t_{0}\right)\left|\left[\mathrm{\hat{H}},\hat{\mathrm{q}}\right]\right|\psi\left(t_{0}\right)\right\rangle =\frac{i}{\hslash}\left\langle \psi\left(t_{0}\right)\left|\left[\mathrm{\hat{H}},\hat{\mathrm{q}}_{H}\right]\right|\psi\left(t_{0}\right)\right\rangle \end{equation}

\begin{equation}
\frac{i}{\hslash}\left\langle \psi\left(t_{0}\right)\left|\left[\mathrm{\hat{H}},\hat{\mathrm{p}}\right]\right|\psi\left(t_{0}\right)\right\rangle =\frac{i}{\hslash}\left\langle \psi\left(t_{0}\right)\left|\left[\mathrm{\hat{H}},\hat{\mathrm{p}}_{H}\right]\right|\psi\left(t_{0}\right)\right\rangle \end{equation}
Now with the help of (12) and (13) and using (1), (8) and (9) can
respectively be rewritten as

\begin{equation}
\dot{Q}=\left\{ Q,\mathscr{H}\right\} ,\end{equation}
and

\begin{equation}
\dot{P}=\left\{ P,\mathscr{H}\right\} ,\end{equation}
where $\mathscr{H}$ denotes the corresponding classical Hamiltonian
due to the Poisson bracket formalism. Equations (14) and (15) are
the classical equations of motion for two variables $P$ and $Q$,
defined by the representations (4) and (5), in an infinitesimal time-interval
when one has neglected terms of $\mathcal{O}\left(\Delta t^{2}\right)$. 

Consequently, when starting out from the Heisenberg picture, terms
of the second-order in time can be neglected, the dynamics of a system
becomes classical. This proves the Feynman postulate and thereby fulfills
the main aim of this Section.

\section{Concluding remark}

The Feynman formalism of quantum mechanics contains a famous postulate
stating that during an infinitesimal time-interval, the propagator
is the classical wave function. This postulate, which was later analyzed
by Holland, was recently highlighted by de Gosson and Hiley by using
an improved version of the Makri-Miller propagator. The question addressed
in the present note was whether the Heisenberg picture of quantum
mechanics can directly handle the Feynman postulate and allow a proof
for that. It was indeed demonstrated that the Heisenberg picture inherently
includes the Feynman postulate and supports a proof for the following
general statement: {}``When terms of $\mathcal{O}\left(\Delta t^{2}\right)$
can be neglected, the dynamics of a system is classical.'' This conclusion
is in agreement with the recent de Gosson-Hiley achievement {[}12{]}---a
result derived from the Bohmian trajectories point of view.

\subsection*{Acknowledgements}

The author is indebted to Nils Schopohl for alerting him to the relation
between quantum mechanics and classical physics.


\begin{thebibliography}{15}
\bibitem[14]{key-24}The coefficient $\hslash$ in (1) was first considered
as an assumption in {[}1{]}, but later, Dirac presented a more rigorous
justification in {[}6{]}.
\bibitem[15]{key-2}He argued convincingly that $h$ is considered
as a fixed number and therefore it is not mathematically sensible
to talk about a limit value for it (e.g., see {[}8{]}).
\end{thebibliography}
\end{document}